\documentclass[conference]{IEEEtran}
\IEEEoverridecommandlockouts
\usepackage{cite}
\usepackage{amsmath,amssymb,amsfonts}
\usepackage{algorithmic}
\usepackage{booktabs,multirow}
\usepackage{colortbl}
\usepackage{graphicx}
\usepackage{makecell}
\usepackage{textcomp}
\usepackage{xcolor}
\def\BibTeX{{\rm B\kern-.05em{\sc i\kern-.025em b}\kern-.08em
    T\kern-.1667em\lower.7ex\hbox{E}\kern-.125emX}}

\usepackage{booktabs}
\usepackage[table]{xcolor}

\begin{document}

\title{Reshaping Checkpoint-Induced Power Variations Through Inference Workload Scheduling
\\
\thanks{This research is sponsored in part by M.C. Dean.}
}

\author{
\IEEEauthorblockN{
Yu Ma, Ziheng Sheng, Anna Andriiko, Chau-Wai Wong,  \\
Jong Ha Woo, Zishuo Yang, Aaditya Pandey, \\
Wenyuan Tang, Ning Lu
}
\IEEEauthorblockA{
\textit{North Carolina State University}\\
Raleigh, NC 27606, USA\\
\{yma47, zsheng2, cwong9, jwoo6, \\zyang66, apande26, wtang8, nlu2\}@ncsu.edu
}
\and
\IEEEauthorblockN{
Andrew Meyers, George Navarro
}
\IEEEauthorblockA{
\textit{M.C. Dean}\\
Tysons, VA 22102, USA\\
\{andrew.meyers, george.navarro\}@mcdean.com
}
}

\maketitle

\begin{abstract}
Synchronous checkpoints in large language model training cause rapid power variations that challenge frequency regulation in islanded data-center microgrids. This paper proposes a checkpoint-aware two-layer framework that schedules flexible inference workloads to compensate for checkpoint-induced disturbances. Using measured GPU power profiles and anticipated checkpoint power deficits, the upper layer coordinates heterogeneous task selection and launch times, while the lower layer adjusts the batch sizes of newly launched tasks online to correct power-tracking errors. Early launching (EL) and staged retirement (SR) further smooth power transitions at checkpoint onset and recovery, respectively. PSCAD/EMTDC simulations with reference checkpoint drops of 1–3 MW demonstrate reductions of 41.6–61.6\% in maximum absolute frequency deviation and 68.5–89.6\% in mean absolute frequency deviation relative to no scheduling. Battery energy storage system (BESS) peak charging power and charging energy decrease by 65.1--100\% and 89.7--100\%, respectively. These benefits exceed those of greedy scheduling when
GPU resources are relatively abundant.
\end{abstract}


\section{Introduction}
Large-scale AI training introduces rapid, workload-dependent power variations that challenge power-system operation \cite{NERC2025LargeLoads}. During large language model (LLM) training, synchronous checkpoints periodically save model states for failure recovery, temporarily interrupting GPU computation and causing sharp power drops followed by rapid restoration. Measurements from a 2,336-A100 GPU cluster show power decreasing from approximately 0.45 MW to below 0.3 MW~\cite{hu2024acmetrace}. When checkpointing is synchronized across large GPU clusters, the resulting aggregated power variations can become substantial, with power transitions occurring much faster than the checkpoint duration.

Such rapid, recurring load variations challenge islanded data-center microgrids with limited generation capacity and inertia. Conventional generators cannot instantly follow abrupt load changes, requiring fast-response resources to maintain power balance. Unlike unexpected load fluctuations, checkpoint events can often be anticipated, enabling proactive scheduling of flexible inference workloads to compensate for checkpoint-induced reductions in power draw.

In computing, CheckFreq \cite{mohan2021checkfreq} reduces checkpoint overhead through adaptive checkpoint frequency and pipelining, while DataStates-LLM \cite{maurya2024datastates} overlaps checkpoint I/O with
computation to reduce interruptions. These approaches primarily target computational efficiency rather than grid-facing power shaping. In power systems, BESSs and supercapacitors mitigate rapid load variations and support frequency stability \cite{aibess2026,aiquality2026}. However, deploying these storage resources requires additional capital investment, while repeated battery cycling may accelerate degradation. Alternatively, workload orchestration supports grid-requested demand reduction \cite{colangelo2026gridinteractive}, while inference batch-size control provides voltage regulation \cite{liang2026gputogrid}. However, these approaches do not explicitly exploit predictable checkpoint timing and training-power deficits for intra-data-center coordination between training and inference workloads. Such checkpoint-aware coordination for mitigating power variations and supporting frequency stability in islanded microgrids remains insufficiently explored.

To address this gap, this paper proposes a checkpoint-aware
hierarchical workload scheduling framework that exploits
measured inference-power profiles to mitigate training
checkpoint disturbances. The main contributions are twofold.
First, an upper-layer scheduler coordinates heterogeneous
inference tasks and optimizes their launch times to
compensate for predicted checkpoint-induced power deficits.
Second, a lower-layer online batch-size adjustment, early launching
and staged task retirement mitigate power-tracking errors
and abrupt power transitions at checkpoint onset and
recovery. PSCAD/EMTDC simulations demonstrate improved
microgrid frequency response and reduced BESS charging
requirements.

\section{Methodology}
\begin{figure*}[t]
    \centering
    \includegraphics[width=\textwidth]{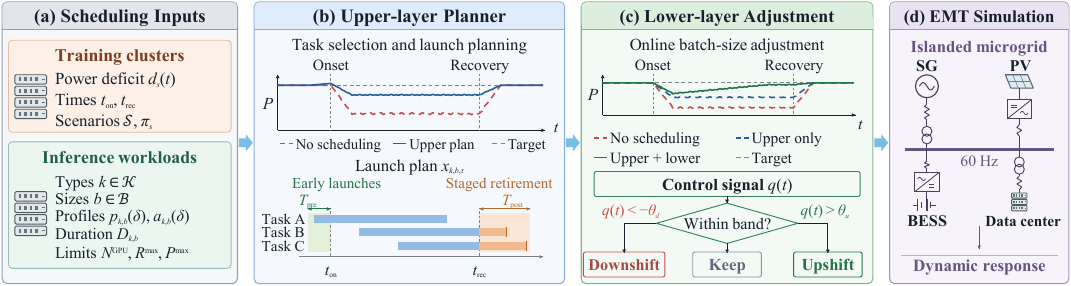}
    \setlength{\abovecaptionskip}{-10pt}
    \caption{Overview of the proposed checkpoint-aware
    hierarchical workload scheduling framework.
    Proactive planning, online batch-size adjustment, and
    checkpoint-boundary shaping coordinate inference demand
    to smooth data-center power, with frequency and BESS
    responses evaluated through EMT simulation.}
    \label{fig:framework}
    \vspace{-10pt}
\end{figure*}

We consider an AI data center with multiple LLM training
clusters and inference clusters serving online and
latency-tolerant workloads. Low GPU utilization during
normal online serving leaves capacity for flexible tasks
whose incremental demand compensates for checkpoint-induced
training-power deficits. As illustrated in Fig.~\ref{fig:framework}, the upper layer
selects tasks and optimizes launch times using predicted
checkpoint power deficits, while the lower layer adjusts
the batch sizes of newly launched tasks to correct
power-tracking errors. Early launching (EL) and staged
retirement (SR) smooth power transitions at checkpoint
onset and recovery, respectively.

\subsection{Workload and Checkpoint Representation}

Let $\mathcal K$ and $\mathcal B$ denote the sets of inference-workload types and batch sizes, respectively. Let $\mathcal T$ denote the discrete optimization horizon and $\mathcal T_{\mathrm L}\subseteq\mathcal T$ the set of candidate launch times. For each workload--batch pair $(k,b)$, measured GPU traces yield an incremental power template $p_{k,b}(\delta)$ relative to the normal light-load online-serving power $p_{\mathrm{GPU}}^{\mathrm{light}}$, where $\delta$ is the elapsed time after launch. These traces also provide the nominal execution duration $D_{k,b}$ and GPU occupancy profile $a_{k,b}(\delta)$, defined as the empirical probability that a task remains active at time $\delta$ after launch.

Checkpoint variability for a given training-cluster setting
is represented by scenarios $s\in\mathcal S$, with
probabilities $\pi_s\ge0$ satisfying
$\sum_{s\in\mathcal S}\pi_s=1$. Each scenario specifies a
checkpoint-induced power deficit $d_s(t)$ describing the
training-power reduction and recovery.

A common launch plan is optimized across these scenarios. Let $x_{k,b,\tau}\in\mathbb Z_{\ge0}$ denote the number of type-$k$, batch-$b$ tasks launched at time $\tau$. Using the nominal power templates, the predicted aggregate incremental inference power is
\begin{equation}
\Delta P_{\mathrm{inf}}(t)=
\sum_{k\in\mathcal K}
\sum_{b\in\mathcal B}
\sum_{\tau\in\mathcal T_{\mathrm L}}
x_{k,b,\tau}p_{k,b}(t-\tau),
\qquad t\in\mathcal T.
\label{eq:inf_power}
\end{equation}
This predicted inference power is scenario-independent,
since the launch plan and templates are shared across
all scenarios. The scheduled data-center power in
scenario $s$ is therefore
\begin{equation}
P_s^{\mathrm{sch}}(t)
=P^{\mathrm{base}}-d_s(t)+\Delta P_{\mathrm{inf}}(t),
\label{eq:dc_power}
\end{equation}
where $P^{\mathrm{base}}$ is the pre-checkpoint
data-center power, including the light-load
online-serving demand of inference GPUs.

\subsection{Checkpoint-Aware Hierarchical Workload Scheduling}
\subsubsection{Upper-Layer Workload Scheduling}

Let $t_{\mathrm{on}}$ and $t_{\mathrm{rec}}$ denote the nominal checkpoint onset
and recovery times. The upper-layer scheduler selects
inference tasks and their launch times over a planning
horizon from $t_{\mathrm{on}}-T_{\mathrm{pre}}$ to
$t_{\mathrm{rec}}+T_{\mathrm{post}}$, where $T_{\mathrm{pre}}$ enables
anticipatory launches and $T_{\mathrm{post}}$ captures
post-checkpoint power transitions.

For scenario $s$, nonnegative variables $u_s(t)$ and
$o_s(t)$ denote power underfill and overshoot, respectively:
\begin{equation}
\begin{aligned}
u_s(t)&\ge d_s(t)-\Delta P_{\mathrm{inf}}(t),\\
o_s(t)&\ge\Delta P_{\mathrm{inf}}(t)-d_s(t).
\end{aligned}
\label{eq:tracking_slacks}
\end{equation}

The upper layer minimizes expected tracking error and
penalizes the maximum power change near checkpoint onset:
\begin{equation}
\min \sum_{s\in\mathcal S}\pi_s
\left[
\sum_{t\in\mathcal T}
\bigl(w^u u_s(t)+w^o o_s(t)\bigr)\Delta\tau
+\lambda_r\mathcal R_s
\right].
\label{eq:upper_objective}
\end{equation}
where
\begin{equation}
\mathcal R_s=
\max_{t\in\mathcal T\cap
[t_{\mathrm{on}}-T_{\mathrm{pre}},\,
 t_{\mathrm{on}}+T_{\mathrm{entry}}]}
\left|P_s^{\mathrm{sch}}(t)
-P_s^{\mathrm{sch}}(t-T_r)\right|.
\label{eq:onset_ramp}
\end{equation}
Here, $w^u$ and $w^o$ are nonnegative weights for underfill and overshoot, respectively, $\Delta\tau$ is the planning time step, and $\lambda_r\ge0$ weights the power-change penalty. $T_{\mathrm{entry}}$ specifies the length of the post-onset evaluation window, while $T_r$ is the time interval over which power changes are measured. Thus, $\mathcal R_s$ penalizes the largest aggregate power change during early launching and initial checkpoint execution.

The schedule is subject to power, GPU occupancy,
and task-launch-rate constraints:
\begin{equation}
\begin{gathered}
P_s^{\mathrm{sch}}(t)\le P^{\max},\ \forall s,t,
\qquad
\sum_{k,b}x_{k,b,\tau}\le R^{\max},\ \forall\tau,\\
\sum_{k,b,\tau}x_{k,b,\tau}a_{k,b}(t-\tau)
\le N^{\mathrm{GPU}},\qquad\forall t.
\end{gathered}
\label{eq:scheduling_constraints}
\end{equation}
Here, $P^{\max}$ is the data-center power limit, $N^{\mathrm{GPU}}$ is the GPU capacity available without violating online-serving requirements, and $R^{\max}$ is the maximum number of tasks launched at each candidate time. Each task occupies one GPU. GPU occupancy is constrained in expectation, with actual availability checked before launch. Resource-infeasible launches are canceled rather than delayed. Candidate tasks must satisfy nominal completion-time requirements.

\subsubsection{Lower-Layer Online Adjustment}

The lower layer corrects deviations between the
upper-layer predicted power and measured data-center
demand by adjusting the batch sizes of newly launched
inference tasks. The tracking error and control signal
are defined as
\begin{equation}
\begin{aligned}
e(t)&=P^{\mathrm{ref}}(t)-P^{\mathrm{meas}}(t),\\
\bar e(t)&=\rho\bar e(t-\Delta t)+(1-\rho)e(t),\\
q(t)&=\alpha e(t)+(1-\alpha)\bar e(t),
\end{aligned}
\label{eq:tracking_error}
\end{equation}
where $P^{\mathrm{ref}}$ and $P^{\mathrm{meas}}$ denote upper-layer predicted and measured data-center power, respectively, $\rho=e^{-\Delta t/T_e}$, and $\alpha$ balances instantaneous and filtered errors.

At each scheduled launch time $\tau_j$, the controller
compares $q_j=q(\tau_j^-)$ with the thresholds
$\theta_u$ and $\theta_d$. An error exceeding
$\theta_u$ triggers a batch-size upshift, while an
error below $-\theta_d$ triggers a downshift.
The number of tasks adjusted is
\begin{equation}
m_j=
\min\left\{
\tilde n_j,\,
\max\left[
1,\operatorname{round}
\left(\frac{|q_j|-\theta_j}{\Delta\bar p_j}\right)
\right]
\right\},
\label{eq:lower_adjust}
\end{equation}
where $\tilde n_j$ is the number of feasible tasks
scheduled for launch, $\theta_j$ is the corresponding
threshold, and $\Delta\bar p_j$ is the expected
short-horizon per-GPU power change estimated from
measured templates of adjacent batch sizes.
No error-triggered adjustment occurs within the
deadband.

Batch-size adjustments apply only to newly launched tasks
and satisfy GPU availability, power-headroom, and
completion-time constraints. During checkpoint recovery,
upshifts are disabled, and staged retirement of preemptible
tasks limits aggregate power overshoot.

\subsubsection{Checkpoint-Boundary Power Shaping}

Valley filling alone does not guarantee smooth power
transitions at checkpoint onset and recovery. At onset,
the gradual power ramp-up of inference workloads limits
their ability to immediately compensate for the
training-power drop. To mitigate this effect, early
launching (EL) allows selected tasks to start within
$[t_{\mathrm{on}}-T_{\mathrm{pre}},t_{\mathrm{on}}]$. Their launch times are
optimized to compensate for the predicted training-power
reduction while limiting aggregate power changes near
checkpoint onset.

At recovery, training-power restoration may cause
excessive aggregate demand if scheduled inference tasks
remain active. Threshold-triggered staged retirement (SR)
progressively terminates selected preemptible tasks when
measured data-center power reaches $P^{\max}$.
Retired GPUs return to their light-load online-serving
state, removing only the incremental power associated
with scheduled tasks. The post-checkpoint window
$[t_{\mathrm{rec}},\,t_{\mathrm{rec}}+T_{\mathrm{post}}]$ captures the resulting
power transitions.

\subsection{Performance Evaluation Metrics}

The proposed framework is evaluated in terms of both microgrid performance and computing-side overhead. Microgrid metrics include BESS charging energy (CE), peak charging power (PCP), and maximum and mean absolute frequency deviations (Max AF and Mean AF). Computing-side costs are assessed through online inference execution time, request termination rate, and wasted request processing time ratio (WRPTR). The WRPTR quantifies the fraction of processing time lost when inference requests are terminated during checkpoint recovery:
\begin{equation}
R_{\mathrm{waste}}=
\frac{\sum_{r\in\mathcal R_{\mathrm{terminated}}}
T_r^{\mathrm{wasted}}}
{\sum_{r\in\mathcal R_{\mathrm{processed}}}
T_r^{\mathrm{processed}}}\times100\%,
\label{eq:waste_ratio}
\end{equation}
where $\mathcal R_{\mathrm{processed}}$ and
$\mathcal R_{\mathrm{terminated}}$ denote started
and forcibly terminated requests, respectively.
$T_r^{\mathrm{processed}}$ excludes queueing time,
and $T_r^{\mathrm{wasted}}$ denotes the non-reusable
processing time before termination.

\section{Simulation Results}
\begin{figure*}[t]
    \centering
    \includegraphics[width=\textwidth]{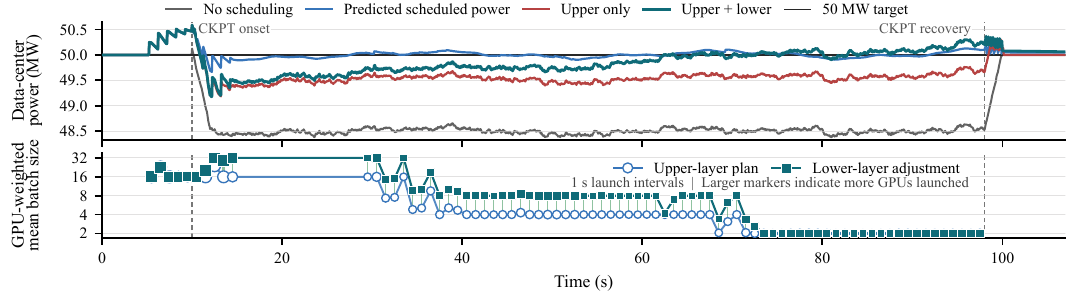}
    \setlength{\abovecaptionskip}{-15pt}
    \caption{Coordination of upper-layer planning and lower-layer
    adjustment. When the checkpoint drop is underestimated,
    online batch-size adjustment brings realized data-center
    power closer to the 50-MW target.}
    \label{fig:performance}
    \vspace{-10pt}
\end{figure*}

A 60-Hz islanded microgrid comprising a synchronous
generator (SG), photovoltaic (PV) generation, a BESS,
and a data center is modeled in PSCAD/EMTDC. The SG
provides primary frequency control, while PV output
is constant. The BESS uses frequency-droop control
with a $\pm0.05$-Hz deadband, an 8-MW/Hz gain, and
$\pm5$-MW power limits. Key parameters are listed
in Table~\ref{tab:simulation_settings}.

\begin{table}[t]
\centering
\caption{Microgrid and Scheduling Parameters}
\label{tab:simulation_settings}
\footnotesize
\renewcommand{\arraystretch}{0.95}
\setlength{\tabcolsep}{3pt}
\begin{tabular*}{\columnwidth}
{@{\extracolsep{\fill}}lcc@{}}
\toprule
Parameter & Symbol & Value \\
\midrule
\rowcolor[gray]{0.92}
\multicolumn{3}{l}{\textbf{Microgrid Parameters}} \\
\addlinespace[2pt]
SG rated capacity   & -- & 50 MVA \\
SG inertia constant & -- & 6 s \\
SG/PV pre-event output & -- & 40 / 10 MW \\
BESS maximum power  & -- & $\pm5$ MW \\
BESS droop gain     & -- & 8 MW/Hz \\
BESS deadband       & -- & $\pm0.05$ Hz \\
\midrule
\rowcolor[gray]{0.92}
\multicolumn{3}{l}{\textbf{Scheduling Parameters}} \\
\addlinespace[2pt]
Schedulable inference GPUs
& $N^{\mathrm{GPU}}$ & 16,304 \\
Launch limit
& $R^{\max}$ & 2,048 \\
Light-load GPU power
& $p_{\mathrm{GPU}}^{\mathrm{light}}$
& 90 W \\
Planning/control intervals
& $\Delta\tau/\Delta t$ & 1 / 0.05 s \\
Tracking weights
& $w^u/w^o$ & 5 / 0.5 \\
Current-error weight
& $\alpha$ & 0.65 \\
Filtering time constant
& $T_e$ & 5 s \\
Upshift threshold
& $\theta_u$ & 0.05 MW \\
Downshift threshold
& $\theta_d$ & 0.0125 MW \\
Advance/post windows
& $T_{\mathrm{pre}}/T_{\mathrm{post}}$ & 5 / 10 s \\
Power-change settings
& $\lambda_r/T_r$ & 100 / 1 s \\
Post-onset evaluation window
& $\mathcal T_{\mathrm{entry}}$ & 10 s \\
Data-center power limit
& $P^{\max}$ & 50.5 MW \\
\bottomrule
\end{tabular*}
\vspace{-12.5pt}
\end{table}

GPU power is measured using \texttt{nvidia-smi} at nominal 100-ms intervals on an NVIDIA H100 running GPT-OSS-20B using Hugging Face Transformers with a PyTorch backend. Inference requests from GSM8K, CommonsenseQA, HotpotQA, and MATH-500 are processed with batch sizes $b\in\{2,4,8,16,32\}$. For each workload--batch-size pair $(k,b)$, traces are split 60:40 for training and testing. The incremental power template $p_{k,b}(\delta)$ is obtained by averaging launch-aligned training traces after subtracting $p_{\mathrm{GPU}}^{\mathrm{light}}$.

Checkpoint scenario sets (A, B, and C) have reference drops and durations of 1 MW/90 s, 2 MW/120 s, and 3 MW/150 s, respectively. Each set contains 50 scenarios, with magnitudes and durations uniformly sampled within $\pm0.5$ MW and $\pm10$ s of their reference values. The first 45 are equally weighted for upper-layer planning, while five are held out for testing. Checkpoint onset and recovery follow 2-s linear ramps. Correlated power noise (0.04-MW standard deviation, 0.8-s time constant) is applied only during checkpointing.

Five baselines are considered: (1) No Scheduling (NS), without checkpoint-aware workload reassignment; (2) Basic Hierarchical Scheduling (BHS), without early launching or staged retirement; (3) BHS+EL, incorporating early launching; (4) BHS+SR, incorporating staged retirement; and (5) Greedy Scheduling (GS), which schedules tasks chronologically by repeatedly selecting feasible tasks that most reduce \eqref{eq:upper_objective}, subject to the launch limit. All scheduled methods employ the same lower-layer batch-size adjustment, while GS and the proposed method incorporate both EL and SR.

\subsection{Performance and Sensitivity Analysis of Hierarchical Workload Scheduling}

Figure~\ref{fig:performance} illustrates the coordination
between upper-layer planning and lower-layer adjustment
for a scenario-A test case with an actual 1.5-MW
checkpoint drop. Although the upper-layer predicted
power remains close to the 50-MW target, the
underestimated checkpoint drop results in a persistent
power deficit under upper-layer scheduling alone.
The lower layer increases the batch sizes of newly
launched inference tasks to compensate for this deficit,
bringing actual data-center power closer to the target.
As the tracking error decreases, the magnitude of
batch-size adjustments is gradually reduced.

Table~\ref{tab:microgrid_results} summarizes the
corresponding microgrid responses. Without scheduling,
the checkpoint-induced load reduction causes a
frequency rise and BESS charging. Upper-layer scheduling
eliminates BESS charging in this test case, while
lower-layer adjustment further reduces the maximum
and mean absolute frequency deviations by 9.1\%
and 48.6\%, respectively, compared with upper-layer
scheduling alone.

\begin{figure}[h]
    \centering
    \includegraphics[width=\columnwidth]{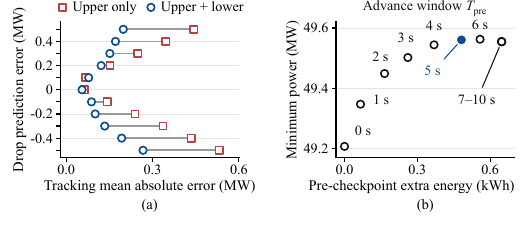}
    \setlength{\abovecaptionskip}{-17.5pt}
    \caption{Sensitivity to (a)~checkpoint prediction error and
    (b)~advance window. Online adjustment reduces tracking errors,
    while a 5-s window balances initial power support and
    pre-checkpoint overshoot.}
    \label{fig:sensitivity}
    \vspace{-10 pt}
\end{figure}

Figure~\ref{fig:sensitivity}(a) evaluates sensitivity
to checkpoint power-prediction errors. Lower-layer
adjustment reduces power-tracking MAE by approximately
49\% and 55\% at prediction errors of $-0.5$ and
$+0.5$ MW, respectively, demonstrating its ability
to mitigate checkpoint prediction uncertainty.
Figure~\ref{fig:sensitivity}(b) examines the effect of the advance window $T_{\mathrm{pre}}$ on checkpoint-entry power shaping. Increasing $T_{\mathrm{pre}}$ improves the minimum data-center power, but the benefit becomes marginal beyond approximately 5 s, while pre-checkpoint extra energy increases. Accordingly, $T_{\mathrm{pre}}=5$ s is selected to balance power compensation and anticipatory energy consumption. These results highlight the trade-off between
checkpoint-entry power smoothing and anticipatory
inference energy consumption.

\begin{table}[t]
\centering
\caption{Microgrid Responses Under Different
Scheduling Strategies}
\label{tab:microgrid_results}
\scriptsize
\setlength{\tabcolsep}{3pt}
\renewcommand{\arraystretch}{1.05}
\begin{tabular}{@{}lcccc@{}}
\hline
Strategy
& \shortstack{CE (kWh)}
& \shortstack{PCP (MW)}
& \shortstack{Max AF (Hz)}
& \shortstack{Mean AF (Hz)} \\
\hline
No scheduling
& 5.339 & 0.5595 & 0.11810 & 0.07635 \\
Upper only
& 0 & 0 & 0.06844 & 0.03049 \\
\textbf{Upper + lower (proposed)}
& \textbf{0} & \textbf{0} & \textbf{0.06222} & \textbf{0.01567} \\
\hline
\end{tabular}
\vspace{-12.5pt}
\end{table}

\subsection{Comparative Evaluation of Scheduling Strategies}

\begin{figure}[h]
    \centering
    \includegraphics[width=\columnwidth]{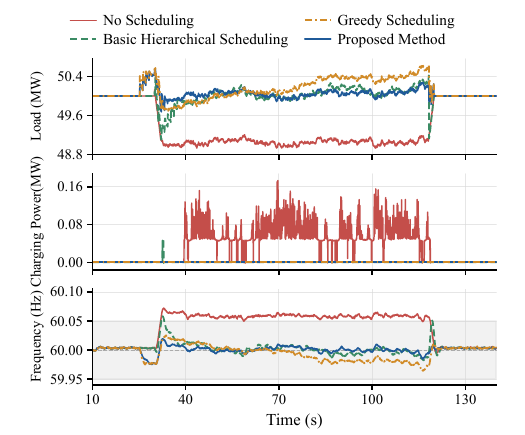}
    \setlength{\abovecaptionskip}{-15pt}
    \caption{Data-center power, BESS charging power, and frequency responses under different scheduling strategies.}
    \label{fig:strategy_response}
    \vspace{-10pt}
\end{figure}

Figure~\ref{fig:strategy_response} compares data-center power, BESS charging, and frequency responses under different scheduling strategies for a scenario-A test case. Compared with no scheduling, BHS substantially reduces the checkpoint-induced power deficit, although power transitions at checkpoint onset and recovery still cause frequency excursions and transient BESS charging. The proposed method further improves power tracking and frequency response through EL and SR.

Table~\ref{tab:drop_comparison} evaluates the contributions of EL and SR across three checkpoint scenario sets. Adding EL to BHS eliminates BESS charging in scenario-A and reduces CE and PCP in scenario-B and -C. In contrast, SR primarily improves recovery-stage behavior, with limited effects on the maximum absolute frequency deviation. Combining both mechanisms provides improvements, demonstrating their complementary roles in checkpoint-boundary power shaping.

\begin{table}[!t]
\centering
\caption{Scheduling Performance Across Checkpoint Scenarios}
\label{tab:drop_comparison}
\scriptsize
\setlength{\tabcolsep}{0.5pt}
\renewcommand{\arraystretch}{0.90}
\begin{tabular*}{\columnwidth}
{@{\extracolsep{\fill}}cccccc@{}}
\toprule
\makecell{Scenario Set}
& \makecell{Strategy}
& \makecell{CE\\(kWh)}
& \makecell{PCP\\(MW)}
& \makecell{Max AF\\(Hz)}
& \makecell{Mean AF\\(Hz)} \\
\midrule
\multirow{6}{*}{A}
& NS & 1.4090 & 0.1762 & 0.0869 & 0.0548 \\
& BHS & 0.0059 & 0.1209 & 0.0660 & 0.0083 \\
& BHS+EL & 0.0000 & 0.0000 & 0.0595 & 0.0070 \\
& BHS+SR & 0.0059 & 0.1210 & 0.0660 & 0.0071 \\
& GS & 0.0000 & 0.0000 & 0.0467 & 0.0137 \\
& \textbf{Proposed}      & \textbf{0.0000} & \textbf{0.0000} & \textbf{0.0334} & \textbf{0.0057} \\
\midrule
\multirow{6}{*}{B}
& NS & 12.282 & 0.9669 & 0.1546 & 0.0971 \\
& BHS & 0.2938 & 0.4472 & 0.1064 & 0.0219 \\
& BHS+EL & 0.1452 & 0.3687 & 0.1052 & 0.0214 \\
& BHS+SR & 0.2430 & 0.4372 & 0.1064 & 0.0209 \\
& GS & 0.2080 & 0.4120 & 0.1015 & 0.0233 \\
& \textbf{Proposed}      & \textbf{0.0944}  & \textbf{0.2350} &\textbf{ 0.0903} & \textbf{0.0202} \\
\midrule
\multirow{6}{*}{C}
& NS & 28.191 & 1.6739 & 0.2264 & 0.1376 \\
& BHS & 3.4411 & 0.7376 & 0.1472 & 0.0451 \\
& BHS+EL & 3.0140 & 0.6322 & 0.1312 & 0.0443 \\
& BHS+SR & 3.3211 & 0.7376 & 0.1472 & 0.0443 \\
& GS & \textbf{2.8520} & 0.6128 & 0.1381 & \textbf{0.0430} \\
& \textbf{Proposed}      & 2.8933 & \textbf{0.5848} & \textbf{0.1303} & 0.0433 \\
\bottomrule
\end{tabular*}
\par\vspace{2pt}
\parbox{\columnwidth}{\scriptsize\raggedright
\textit{Note:} Means over five random test scenarios per set.
}
\vspace{-15pt}
\end{table}

Compared with GS, the proposed method improves frequency response in scenario sets A and B. Fig.~\ref{fig:strategy_response} shows that GS overfills the power valley during the later checkpoint period, resulting in frequency undershoot. By considering power-tracking requirements and GPU availability over the planning horizon, the proposed method reduces mismatches. In scenario-C, limited GPU capacity constrains both methods to nearly identical scheduling decisions. Consequently, their system dynamic responses and performance metrics are similar, reflecting the limited flexibility under resource saturation.

\subsection{Computing-Side Costs of Workload Scheduling}

GPU reassignment for checkpoint compensation may require higher concurrency on the remaining GPUs to serve online requests. Table~\ref{tab:request_completion} evaluates the associated completion time costs. Completion times remain nearly unchanged for CommonsenseQA and GSM8K across the tested batch sizes, while HotpotQA and MATH-500 exhibit substantial increases in mean completion time at $b=16$ and $32$. At $b=32$, their mean completion times increase by 121.2\% and 199.0\%, respectively, compared with $b=2$. The large mean-to-median differences further indicate considerable completion-time variability at high batch sizes, highlighting the need for workload-specific concurrency limits.

SR also introduces computational overhead by terminating active inference requests during checkpoint recovery. Across scenario sets A, B, and C, the request termination rates are 4.8\%, 3.7\%, and 3.5\%, respectively, with corresponding WRPTRs of 7.4\%, 5.8\%, and 4.5\%. These results reveal a trade-off between checkpoint-recovery power smoothing and the computational work lost through task termination.

\begin{table}[t]
\centering
\caption{Request Completion Times under Different Batch Sizes}
\label{tab:request_completion}
\scriptsize
\setlength{\tabcolsep}{1.2pt}
\renewcommand{\arraystretch}{0.9}
\begin{tabular*}{\columnwidth}
{@{\extracolsep{\fill}}lccccc@{}}
\toprule
Dataset & $b=2$ & $b=4$ & $b=8$ & $b=16$ & $b=32$ \\
\midrule
CommonsenseQA
& 22.0 / 12.9 & 22.1 / 13.4 & 22.2 / 13.0
& 21.4 / 13.0 & 22.3 / 13.6 \\
GSM8K
& 20.6 / 15.8 & 20.9 / 15.9 & 20.7 / 16.4
& 20.3 / 15.8 & 22.1 / 17.6 \\
HotpotQA
& 77.7 / 22.0 & 77.8 / 21.9 & 77.8 / 23.6
& 131.5 / 37.3 & 171.9 / 25.2 \\
MATH-500
& 90.3 / 32.1 & 91.7 / 35.5 & 94.1 / 36.1
& 141.1 / 41.3 & 269.9 / 41.6 \\
\bottomrule
\end{tabular*}
\par\vspace{2pt}
\parbox{\columnwidth}{\scriptsize\raggedright
\textit{Note:} Entries report mean/median
request completion times in seconds.}
\vspace{-17.5pt}
\end{table}

\section{Conclusion}


This paper proposes a checkpoint-aware hierarchical inference scheduling framework to mitigate power variations in islanded data-center microgrids. Upper-layer planning, online batch-size adjustment, early launching, and staged retirement jointly reduce frequency deviations and BESS charging requirements. Results also reveal computing-side trade-offs, including increased completion times and wasted processing. Future work will consider dynamic request arrivals and joint workload–BESS optimization.

\color{black}

\color{black}

\bibliographystyle{IEEEtran}
\bibliography{references}
\end{document}